# Controlling for individual heterogeneity in longitudinal models, with applications to student achievement*

**J.R. Lockwood and Daniel F. McCaffrey**

*RAND*
*4570 Fifth Avenue, Suite 600*
*Pittsburgh, PA 15213*
*e-mail:* lockwood@rand.org*;* danielm@rand.org

**Abstract:**
Longitudinal data tracking repeated measurements on individuals are highly valued for research because they offer controls for unmeasured individual heterogeneity that might otherwise bias results. Random effects or mixed models approaches, which treat individual heterogeneity as part of the model error term and use generalized least squares to estimate model parameters, are often criticized because correlation between unobserved individual effects and other model variables can lead to biased and inconsistent parameter estimates. Starting with an examination of the relationship between random effects and fixed effects estimators in the standard unobserved effects model, this article demonstrates through analysis and simulation that the mixed model approach has a "bias compression" property under a general model for individual heterogeneity that can mitigate bias due to uncontrolled differences among individuals. The general model is motivated by the complexities of longitudinal student achievement measures, but the results have broad applicability to longitudinal modeling.

**Keywords and phrases:** longitudinal data analysis, random effects models, fixed effects models, omitted variables.

## Contents

*This material is based on work supported by the Department of Education Institute of Education Sciences under Grant No. R305U040005 and by RAND. Any opinions, findings and conclusions or recommendations expressed in this material are those of the authors and do not necessarily reflect the views of these organizations. We thank Tim Sass, Anastasia Semykina, and S. Paul Wright for helpful comments that substantially improved the manuscript.

## 1. Introduction

Longitudinal data are highly valued in all areas of social science research. In education research in particular, the rapidly growing availability of data tracking student achievement over time has made longitudinal data analysis increasingly prominent. Longitudinal data analysis is now common practice in research on identifying effective teaching practices, measuring the impacts of teacher credentialing and training, and evaluating other educational interventions (8; 14; 15; 19; 20; 26; 42; 52). In fact, the United States Department of Education funded a center dedicated to longitudinal analyses in education (`http://www.caldercenter.org/`). Recent computational advances and empirical findings about the impacts of individual teachers have also intensified interest in "value-added" methods (VAM), where trajectories of students' test scores as they progress through schooling are used to estimate the contributions of individual teachers or schools to student achievement (6; 9; 23; 24; 27; 34; 41). With teacher and school accountability at the forefront of education policy, and with educators and researchers seeking more sophisticated ways of putting test score data to good use, longitudinal methods are likely to remain critical to education research.

One of the most important attributes of longitudinal data for research is that they can lead to less biased and more precise estimates of the effects of substantive variables on individual outcomes than is generally possible with purely cross-sectional observational data. For example, in education research, analysts have consistently found that observable background characteristics of students typically available in administrative databases, such as race/ethnicity, socio-economic status indicators, and limited English proficiency status, are correlated with student achievement, but that there remains substantial unexplained heterogeneity among students in achievement profiles after accounting for these observable characteristics (20). In observational studies with cross-sectional data, this unmeasured heterogeneity threatens to bias estimates of the effects of educational variables being studied (e.g. teacher characteristics such as certification and years experience) because of non-random allocation of students to educational settings such as schools, classrooms, or programs. However, the repeated measures on individuals inherent to longitudinal data provide opportunities to control for this unmeasured heterogeneity, thereby improving the quality of the estimates.

Statisticians and economists often take different approaches to longitudinal data analyses. Common approaches in the statistical literature focus on modeling growth curves and modeling unobserved heterogeneity as part of the error structure for the data using random effects or mixed models (10; 21; 37; 45). Economists have tended to focus more the potential biasing effects of unobserved heterogeneity and fixed effects approaches to remove that bias under the appropriate assumptions. Fixed effects approaches introduce parameters for each individual as part of the model mean structure, rather than the error structure (13; 22; 50). The divide between fixed and random effects approaches is particularly strong in education research, with numerous examples of economists using fixed effects (14; 15; 20; 25; 38; 39; 49; 52), and educational statisticians using random effects or mixed model approaches by way of hierarchical linear models and related methods (6; 26; 30; 29; 32; 33; 36; 37; 40; 41; 53).

The usual criticism of the random effects or mixed model approach is that when treating individual heterogeneity as part of the model error term, correlation between the unobserved individual effects and other substantive variables in the model can lead to biased and inconsistent estimates of the effects of those variables, while fixed effects approaches do not suffer these same shortcomings (3; 13; 16; 20; 22; 50). However, it is also the case that under the standard model of time-invariant individual effects, when many measures are available for each individual, and/or when individual heterogeneity accounts for large fraction of the observed variance in the measurements, the magnitude of the bias from the mixed model approach can be small (13; 22; 50). The goal of this article is to demonstrate that the bias from the mixed model approach can also be small under a model that generalizes the standard unobserved effects model by allowing for multiple time-invariant individual parameters that are related in time-varying ways to the observed measurements. Such a model is motivated by the complexities of longitudinal student achievement data but has broad applicability to other outcome variables in other research areas. Obtaining consistent estimates of parameters under this model would generally require more sophisticated methods and assumptions. Thus, the practical conclusion from this article is that there are circumstances in longitudinal analyses where the mixed model approach can provide reasonable conclusions that are robust across potentially complex structural models for individual heterogeneity, even if that heterogeneity is correlated with the substantive variables being studied.

We begin by defining the standard model with time-invariant individual parameters in Section 2. We compare the fixed and random effects estimators under this model, and briefly review results about the conditions under which the two approaches will lead to similar estimates. In Section 3 we then generalize the standard unobserved effects model to a general mixed model and show how, under mild assumptions, the mixed model approach has a certain "bias compression" property that can mitigate bias due to uncontrolled differences among individuals. We provide simulation examples in Section 4, representative of common empirical scenarios, that demonstrate the implications of our analytical results. We conclude with a discussion of practical considerations and suggestions for future research in Section 5.

## 2. Fixed and Random Effects Estimators in the Standard Case

In this section we consider fixed and random effects estimators under the standard structural model of time-invariant individual effects. We briefly review results about the relationship between the estimators and its dependence on the number of observations for each individual and the strength of their correlation within individuals. We then provide an alternative view of the relationship between the estimators that facilitates comparison to the more complex case considered later in the article. Consistent with our theme of applications to student achievement, we refer to the individuals as students and the measurements as achievement test scores, but all results hold more generally.

### *2.1. Model Specification*

We assume that $T$ achievement measures are tracked for each of $n$ students. The students may represent multiple cohorts but for simplicity of the description we act if is the students belong to a single cohort. The model in this section is most applicable when these achievement measures are taken on a single subject (e.g. reading or mathematics) at different time points, and thus we refer to measurements $t = 1, \ldots, T$ in terms of time. Often $t = 1, \ldots, T$ corresponds to grades but this is not required; for example, some schools and districts administer multiple assessments of the same academic subject during each grade. We let $\boldsymbol{Y}_i = (Y_{i1}, \ldots, Y_{it})$ denote the vector of achievement scores [1] for student $i$ and posit the model

$$\boldsymbol{Y}_i = \boldsymbol{Z}_i\boldsymbol{\theta} + \mathbf{1}\delta_i + \boldsymbol{\epsilon}_i \tag{1}$$

The design matrix $\boldsymbol{Z}_i$ is $(T \times k)$ and has an associated $(k \times 1)$ parameter $\boldsymbol{\theta}$ which is unknown and is the objective of inference. Note that in general, $k$ may be a function of $T$ because of the addition of covariates (e.g. timepoint means) as $T$ grows. Each student has a specific effect $\delta_i$ that applies to all scores via the $(T \times 1)$ vector $\mathbf{1}$. The treatment of this effect as either fixed or random in the process of estimating $\boldsymbol{\theta}$ is the primary consideration of this section. We make no assumption about the relationship of $\delta_i$ to the other covariates in the model, leaving open the possibility that $E(\delta_i|\boldsymbol{Z}_i) \neq 0$. The residual error term $\boldsymbol{\epsilon}_i$ is assumed to be $N(\mathbf{0}, \sigma^2\boldsymbol{I})$, is assumed to be independent of $\delta_i$, and is assumed to satisfy $E(\boldsymbol{\epsilon}_i|\boldsymbol{Z}_i) = \mathbf{0}$ (throughout the article we use $\boldsymbol{I}$ to denote identity matrices of conforming size).

The design matrix $\boldsymbol{Z}_i$ is general. In educational applications, $\boldsymbol{Z}_i$ might typically include time marginal means, time-invariant characteristics of individual teachers, time-varying teacher-level or classroom-level predictors, time-varying or time-invariant (if students switch schools) school factors, or time-varying student characteristics. Importantly, we assume that $\boldsymbol{Z}_i$ does not include time-invariant student characteristics. In that case, the parameters for those characteristics are identified only if the individual student effects are treated as random effects. If such coefficients are of primary interest, then one is forced to use random effects (3; 13; 50) and the comparison of the two estimators is meaningless. We are more interested in considering cases where both approaches are possible.

[1] We use "achievement score" generally, and we allow the possibility that the scores are actually annual gain scores (i.e. $Y_{i,t} - Y_{i,t-1}$) which are commonly used directly as outcome variables in the education research literature. The algebra applied to the model to produce the estimators is equally valid if the $Y_{it}$ were to be annual gain scores rather than level scores, but the assumptions of the model might be more appropriate for level scores than for gain scores.

The expanded version of the model in Equation 1 for the stacked vector of student measurements $\boldsymbol{Y}$ (of length $nT$) across both students and time is

$$\boldsymbol{Y} = \boldsymbol{Z\theta} + \boldsymbol{D\delta} + \boldsymbol{\epsilon} \tag{2}$$

where $\boldsymbol{Z}$ is $(nT \times k)$ obtained by stacking the $\boldsymbol{Z}_i$, $\boldsymbol{D}$ is a $(nT \times n)$ matrix of indicators or dummy variables linking records to the student effects given by the $n$-vector $\boldsymbol{\delta}$, and $\boldsymbol{\epsilon}$ is a $nT$-vector with distribution $N(\boldsymbol{0}, \sigma^2\boldsymbol{I})$, independent of $\boldsymbol{\delta}$, and satisfies $E(\boldsymbol{\epsilon}|\boldsymbol{Z}) = \boldsymbol{0}$.

### 2.2. The Fixed Effects Estimator for $\boldsymbol{\theta}$

The fixed effects estimator is obtained by ordinary least squares, treating each $\delta_i$ as a model parameter. These individual student effects are nuisance. The parameters of interest are $\boldsymbol{\theta}$ and the estimator of these parameters in the presence of the nuisance parameters can be obtained through a two-stage regression. In the first stage, we regress the student fixed effects ($\boldsymbol{D}$) out of both $\boldsymbol{Y}$ and each of the columns of $\boldsymbol{Z}$. $\widehat{\boldsymbol{\theta}}_F$ is then obtained by regression on the resulting sets of residuals.[2] Letting $\boldsymbol{H}_D = \boldsymbol{D}(\boldsymbol{D'D})^{-1}\boldsymbol{D'}$ be the "hat" or projection matrix from the first-stage regression on the student fixed effects, the residuals from the regression of $\boldsymbol{Y}$ on $\boldsymbol{D}$ are given by $(\boldsymbol{I} - \boldsymbol{H}_D)\boldsymbol{Y}$, and similarly the residuals from the regression of $\boldsymbol{Z}$ on $\boldsymbol{D}$ are given by $(\boldsymbol{I} - \boldsymbol{H}_D)\boldsymbol{Z}$. Using the fact that $(\boldsymbol{I} - \boldsymbol{H}_D)$ is symmetric and idempotent, the estimator $\widehat{\boldsymbol{\theta}}_F$ from the second stage regression is

$$\widehat{\boldsymbol{\theta}}_F = \left(\boldsymbol{Z'}(\boldsymbol{I} - \boldsymbol{H}_D)\boldsymbol{Z}\right)^{-1} \boldsymbol{Z'}(\boldsymbol{I} - \boldsymbol{H}_D)\boldsymbol{Y} \tag{3}$$

As will be shown later this form is also convenient for comparing the fixed effects and mixed model estimators.

The transformation of $\boldsymbol{Z}$ and $\boldsymbol{Y}$ by $(\boldsymbol{I} - \boldsymbol{H}_D)$ subtracts from the elements of each vector the within-student averages of those elements. For example, the elements of $(\boldsymbol{I} - \boldsymbol{H}_D)\boldsymbol{Y}$ are $(Y_{it} - \bar{Y}_i)$, where $\bar{Y}_i$ is the average of the scores for the $i$th student. This differencing effectively removes the unobserved heterogeneity, so that under minimal assumptions about $\boldsymbol{Z}$ the fixed effects estimator is consistent for $\boldsymbol{\theta}$ as the number of students gets large (13; 22; 50).

[2] We are implicitly assuming that the matrix $[\boldsymbol{Z}|\boldsymbol{D}]$ is full column rank, which would not be the case if for example $\boldsymbol{Z}$ included indicator columns for all teachers. In that case, columns for some teachers would need to be excluded from $\boldsymbol{Z}$ and the appropriate changes would need to be made to the random effects model to maintain comparability.

### 2.3. The Random Effects Estimator for $\boldsymbol{\theta}$

The random effects estimator treats the student effects $\delta_i$ as random effects with variance $\nu^2$. Conditional on known values of $\nu^2$ and the residual variance $\sigma^2$ and letting $\boldsymbol{R}$ be the covariance matrix of the terms $\boldsymbol{D\delta} + \boldsymbol{\epsilon}$, $\boldsymbol{\theta}$ is estimated by generalized least squares (GLS) via

$$\widehat{\boldsymbol{\theta}}_R = \left(\boldsymbol{Z}'\boldsymbol{R}^{-1}\boldsymbol{Z}\right)^{-1}\boldsymbol{Z}'\boldsymbol{R}^{-1}\boldsymbol{Y} \tag{4}$$

Motivation for the GLS estimator and the standard properties of the estimator are based on the assumption that the $\delta_i$ are independent of the other variables in the model. However, the estimator can be used even if that assumption is violated and the resulting estimator can still have desirable properties, which is one of the main themes of the article. It is also important to note that the estimator in Equation 4 is commonly termed the "infeasible GLS estimator" because it assumes known values of the variance components, which are generally not available. A so-called "feasible GLS estimator" includes a step that uses the data to obtain a consistent estimate $\widehat{\boldsymbol{R}}$ of $\boldsymbol{R}$, and then uses $\widehat{\boldsymbol{R}}$ in Equation 4 (50). We revisit this important distinction in both the Simulation Examples and Discussion Sections.

Based on evaluation of $\boldsymbol{R}^{-1}$, it can be shown that (13; 22; 50)

$$\widehat{\boldsymbol{\theta}}_R = \left(\boldsymbol{Z}'\left(\boldsymbol{I} - \gamma T\boldsymbol{H}_D\right)\boldsymbol{Z}\right)^{-1}\boldsymbol{Z}'\left(\boldsymbol{I} - \gamma T\boldsymbol{H}_D\right)\boldsymbol{Y} \tag{5}$$

where

$$\gamma = \frac{\rho}{1+\rho(T-1)} \tag{6}$$

and $\rho$ is student-level intra-class correlation $\nu^2/(\nu^2+\sigma^2)$.

Equations 5 and 3 show that for this simple model, the fixed effects and mixed models estimators are highly similar, deviating only by the term $\gamma T$. Wooldridge (50) refers to the transformation by $(\boldsymbol{I} - \gamma T\boldsymbol{H}_D)$ as "quasi-time demeaning" because it is equivalent to subtracting a fraction $\gamma T$ of the within-student averages of the components. Moreover, as $\gamma T$ approaches 1 the fixed effects and mixed model estimators will tend to be very similar (13; 22; 50).

Referring to the definition of $\gamma$ in Equation 6, as $\rho \to 1$ for fixed $T$, $\gamma T \to 1$, and as $T \to \infty$ for fixed $\rho > 0$, $\gamma T \to 1$. When $\rho \to 1$ for fixed $T$, the continuity of matrix operations implies that we can write $(\widehat{\boldsymbol{\theta}}_F - \widehat{\boldsymbol{\theta}}_R) = \boldsymbol{QY}$ where the elements of $\boldsymbol{Q}$ are $o(1)$, so that $(\widehat{\boldsymbol{\theta}}_F - \widehat{\boldsymbol{\theta}}_R) \xrightarrow{p} \mathbf{0}$. The analogous result for fixed $\rho$ as $T \to \infty$; namely, that $(\widehat{\boldsymbol{\theta}}_F - \widehat{\boldsymbol{\theta}}_R) \xrightarrow{p} \mathbf{0}$ as $T \to \infty$ for fixed $\rho > 0$, also holds under some additional assumptions on the design matrix $\boldsymbol{Z}$ (28). In practical terms, these results imply that if either many scores are available for each student, or if $\rho$ is large, $\gamma T \approx 1$ and the estimators should be similar across a broad range of $\boldsymbol{Z}$. For example, when $T = 5$ and $\rho$ in the range of 0.7 to 0.8, values typical of actual longitudinal achievement data series, $\gamma T$ is in the range of 0.92 to 0.95.

Under the model in Equation 1, the expected value of the mixed model estimator from Equation 5 conditional on $\boldsymbol{Z}$ is (given $\gamma$) is

$$E(\widehat{\boldsymbol{\theta}}_R|\boldsymbol{Z}) = \boldsymbol{\theta} + (1-\gamma T)\left(\boldsymbol{Z}'\left(\boldsymbol{I}-\gamma T\boldsymbol{H}_D\right)\boldsymbol{Z}\right)^{-1}\boldsymbol{Z}'\boldsymbol{D}E(\boldsymbol{\delta}|\boldsymbol{Z}) \tag{7}$$

which follows from that the fact that $\boldsymbol{H}_D\boldsymbol{D} = \boldsymbol{D}$. The second term on the right hand side is not in general zero when $E(\boldsymbol{\delta}|\boldsymbol{Z}) \neq 0$, and thus the random effects estimator does not guarantee the elimination of the selection bias. Even as $n \to \infty$ the second term on the RHS does not in general tend to zero and thus the random effects estimator is not consistent, the standard result leading many practitioners to prefer fixed effects. However, the leading coefficient of $(1-\gamma T)$ on the bias term in Equation 7 indicates that the random effects estimator "compresses" the bias toward zero, with the degree of compression increasing as $\gamma T \to 1$. This provides an alternative view to quasi-time demeaning of the mechanism by which random effects estimators can mitigate bias. Importantly, this bias compression is a feature of the random effects estimator that holds under more general structural models considered next.

## 3. Extensions to More Complex Models

In this section we consider a generalization of the structural model in Equation 1 that allows for multiple time-invariant individual effects that can be related to the measurements in ways that vary across time. Such a model is particularly relevant for standardized test score data, but is likely applicable to many kinds of outcomes used in social science research. We provide a theorem about the conditions under which the bias compression property of the random effects estimator previously demonstrated for the standard model carries over to mixed effects estimation of the more complex structural model. We also present some illustrative examples.

### *3.1. Model Specification*

The model in Equation 1 assumes that the individual effect $\delta_i$ is related to each measurement in exactly the same way. This assumption is the key to the ability of the fixed effects estimator to remove bias due to individual heterogeneity, because it implies that within-individual differences of measurements do not depend on $\delta_i$. While this assumption may provide adequate approximations in many circumstances, it is unlikely to be exactly met with outcomes commonly encountered in education and other research areas.

For example, the complexities of creating and scaling standardized achievement assessments, including multidimensionality of measured constructs, content shift over time, and vertical equating procedures (11; 17; 29; 31; 43). may make the constant additive effect of Equation 1 too rigid to adequately capture the relationships among multiple scores taken from the same student over time, even if those tests are from a single test developer and are intended to provide measurements on a common scale. For instance, mathematics tests might contain some items related to algebraic concepts and other items related to arithmetic computations, and the proportion of items from each domain might change as students mature from elementary school to secondary school, thus resulting in achievement test scores that differentially combine student achievement on two constructs. Even circumstances as relatively simple as differential reliability of assessments across time may be sufficient to invalidate the structural model in Equation 1. And with criterion-referenced tests (tests designed to measure student performance relative to absolute standards rather than relative to other students) that are not vertically equated becoming more common in response to the requirements of the Federal No Child Left Behind Act, it is likely that many longitudinal achievement data series cannot be assumed to provide measures of a single, consistently scaled construct over time.

We thus consider the following generalization to the model for the vector of scores for one student:

$$\boldsymbol{Y}_i = \boldsymbol{Z}_i\boldsymbol{\theta} + \boldsymbol{A}_1\boldsymbol{\delta}_i + \boldsymbol{\epsilon}_i \tag{8}$$

The assumptions about the design matrix $\boldsymbol{Z}_i$ are unchanged from the standard case in Equation 1. But we generalize the model for student heterogeneity by replacing the scalar $\delta_i$ specific to student $i$ with a $d$-dimensional vector of factors $\boldsymbol{\delta}_i$, and we allow those factors to be represented in the tests as arbitrary linear combinations that can vary over time via the $(T \times d)$ matrix $\boldsymbol{A}_1$[3]. We assume that the factors are mean zero, normally distributed with $V(\boldsymbol{\delta}_i) = \boldsymbol{S}_1$, a $(d \times d)$ positive definite matrix. The arbitrary covariance structure for the factors allows them to cover such cases as different aspects of mathematics ability that might be positively correlated (e.g. problem solving and computation abilities).

By allowing the $\boldsymbol{\delta}_i$ to have multiple components, Equation 8 can account for multi-dimensionality of tests and changing weights on these constructs over measurements. For example, suppose a test measures $d$ constructs so that scores depend on $d$ factors. We let the vector $\boldsymbol{\delta}_i$ denote the time-invariant values on these factors for student $i$. Row $t$ of $\boldsymbol{A}_1$ contains the weights for these factors for the $t$th measurement. As the measures change, the values in the rows of $\boldsymbol{A}_1$ change to allow differential weighting of the factors. Examples 1, 2, and 4 below provide specific examples of this type of scenario for test scores. Random polynomial growth models, considered in Example 3, are also a special case of Equation 8. In this case, $\boldsymbol{\delta}$ contains the random coefficients of the polynomial growth model and the columns of $\boldsymbol{A}_1$ are the polynomials of time. It is important to note that the random growth model is a special case because the values of $\boldsymbol{A}_1$ are assumed to be known, which provided that $T$ is sufficiently large relative to $d$, allows both $\boldsymbol{S}_1$ and thus the individual components of $\boldsymbol{\delta}_i$ to be identified. In general, it will not be possible to separately identify $\boldsymbol{A}_1$ and $\boldsymbol{S}_1$, but that is not problematic because as discussed below the mixed model estimator for $\boldsymbol{\theta}$ does not require this identification.

[3] Block diagonal matrices appear several times in this remainder of this article. When the blocks are equal across individuals we use a bold-face English letter to denote the block diagonal matrix and the same letter with the subscript "1" to denote an individual block. When the blocks differ by student we use the letter with the subscript $i$ to denote the individual blocks.

Another important class of models covered by Equation 8 are those that jointly model measurements from different tested subjects such as mathematics and reading. The multi-factor formulation is perfectly suited to joint longitudinal modeling of outcomes from different tested subjects by treating the models for different subjects as a set of seemingly unrelated regressions (SUR) (51) where the factors can represent different ability attributes relevant for different subjects. In this case it is important to keep in mind that our subscript $t$ does not strictly represent time, but rather more generally indexes repeated measures on students that may be taken both across time and across subjects within time.

The factors are assumed to be independent across students, but as before we allow the possibility that $E(\boldsymbol{\delta}_i|\boldsymbol{Z}) \neq \mathbf{0}$. We further assume that $rank(\boldsymbol{A}_1) = d$ for all $T$, which essentially means that we have chosen a parameterization of the factors $\boldsymbol{\delta}_i$ with minimal dimension. If $rank(\boldsymbol{A}_1) = r < d$, it can be shown that it is possible to reduce the factor to one of dimension $r$ and recover the same marginal covariance structure of the student heterogeneity terms, so without loss of generality we assume that we are dealing with this maximally parsimonious representation at the outset.

The residual error term $\boldsymbol{\epsilon}_i$ is assumed to be mean zero, to be independent of $\boldsymbol{\delta}_i$ for each $i$, to be independent across students, and to satisfy $E(\boldsymbol{\epsilon}_i|\boldsymbol{Z}) = \mathbf{0}$ for each $i$. We let $V(\boldsymbol{\epsilon}_i) = \boldsymbol{\Psi}_1$, a positive definite matrix with diagonal elements bounded away from $\infty$. In many practical cases it may be reasonable to assume that $\boldsymbol{\Psi}_1$ is diagonal, but this restriction is not required. For example, $\boldsymbol{\epsilon}_i$ may have an autoregressive structure. These assumptions imply that $\boldsymbol{R}_1 := V(\boldsymbol{A}_1\boldsymbol{\delta}_i + \boldsymbol{\epsilon}_i) = \boldsymbol{A}_1\boldsymbol{S}_1\boldsymbol{A}_1' + \boldsymbol{\Psi}_1$, the usual form considered in factor analysis when $\boldsymbol{S}_1 = \boldsymbol{I}$ (4; 35). This structural model recovers the standard model considered in Section 2 by taking $d = 1$, $\boldsymbol{S}_1 = \nu^2$, $\boldsymbol{A}_1 = \mathbf{1}$ and $\boldsymbol{\Psi}_1 = \sigma^2\boldsymbol{I}$. However, by allowing the possibility of multiple student-specific factors that link differentially to the test scores, and by allowing different residual variances across time points, this model is capable of expressing arbitrary covariance structures of the student-specific portion of the model. As noted, the generality of this model was motivated by considerations of longitudinal standardized test score data, but is relevant beyond this specific application.

The expanded version of the model for all student measurements

$$\boldsymbol{Y} = \boldsymbol{Z\theta} + \boldsymbol{A\delta} + \boldsymbol{\epsilon} \tag{9}$$

is analogous to the one presented in Equation 2, where again $\boldsymbol{Z}$ is $(nT \times k)$ obtained by stacking the $\boldsymbol{Z}_i$, $\boldsymbol{A} = \boldsymbol{I}_n \otimes \boldsymbol{A}_1$ is $(nT \times nd)$, $\boldsymbol{\delta}$ is length $nd$ obtained by stacking the student factor vectors $\boldsymbol{\delta}_i$, and $\boldsymbol{\epsilon}$ is the vector of $nT$ residual errors with covariance matrix $\boldsymbol{I}_n \otimes \boldsymbol{\Psi}_1$. The covariance matrix of the student portion of the model, $(\boldsymbol{A\delta} + \boldsymbol{\epsilon})$ is thus $\boldsymbol{R} = \boldsymbol{I}_n \otimes \boldsymbol{R}_1 = \boldsymbol{I}_n \otimes (\boldsymbol{A}_1\boldsymbol{S}_1\boldsymbol{A}_1' + \boldsymbol{\Psi}_1)$.

### 3.2. Bias Compression Under the General Model

The mixed model estimator under Equation 9, conditional on the value of $\boldsymbol{R}$, is identical to the random effects estimator of Equation 4, except that $\boldsymbol{R}$ will have more complex structure in the mixed model than it did in the simple random effects model. The mixed model estimator is also the GLS estimator assuming the residual covariance matrix $\boldsymbol{R}$. In some circumstances $\boldsymbol{R}$ may be parameterized and estimated via random effects or random coefficients, for example, in hierarchical linear models (37), but this is not necessary. For example, in value-added modeling of individual teacher effects, several prominent models allow $\boldsymbol{R}$ to be unstructured (41; 33). Later we discuss practical considerations about when it is feasible to impose no structure on $\boldsymbol{R}$ and when parameter-reducing assumptions might be required.

Under the structural model in Equation 9, the expected value of the mixed model estimator (assuming known $\boldsymbol{R}$) is

$$E(\widehat{\boldsymbol{\theta}}_R|\boldsymbol{Z}) = \boldsymbol{\theta} + \left(\boldsymbol{Z}'\boldsymbol{R}^{-1}\boldsymbol{Z}\right)^{-1}\boldsymbol{Z}'\boldsymbol{R}^{-1}\boldsymbol{A}E(\boldsymbol{\delta}|\boldsymbol{Z}) \tag{10}$$

which should be compared to Equation 7 with the matrix $\boldsymbol{R}^{-1}\boldsymbol{A}$ generalizing the matrix $(1-\gamma T)\boldsymbol{D}$. Assuming that the diagonals of $\boldsymbol{\Psi}$ are bounded away from zero, $\boldsymbol{R}^{-1}$ is positive definite which implies that $\boldsymbol{R}^{-1}\boldsymbol{A} = \boldsymbol{0}$ only in the degenerate case of $\boldsymbol{A} \equiv \boldsymbol{0}$. Thus, in general the estimator is neither unbiased nor consistent as $n \to \infty$ when $E(\boldsymbol{\delta}|\boldsymbol{Z}) \neq 0$. However, the elements of $\boldsymbol{R}^{-1}\boldsymbol{A}$ can be very close to 0. In Appendix A we prove the following theorem, which implies that in many practically relevant cases, the elements of $\boldsymbol{R}^{-1}\boldsymbol{A}$ approach zero as $T \to \infty$:

**Theorem.** *Let $\boldsymbol{A}_1$ and $\boldsymbol{\Psi}_1$ be defined as above. Then sufficient conditions for the elements of $\boldsymbol{R}^{-1}\boldsymbol{A}$ to go to zero uniformly as $T \to \infty$ are:*

1. *The smallest eigenvalue of $\boldsymbol{A}_1'\boldsymbol{\Psi}_1^{-1}\boldsymbol{A}_1$ goes to infinity as $T \to \infty$; and*
2. *There exists a number $C$ independent of $T$ such that the elements $a_{it}$ of $\boldsymbol{\Psi}_1^{-1/2}$, the symmetric square root of $\boldsymbol{\Psi}_1^{-1}$, satisfy $\sum_{t=1}^{T}|a_{it}| < C$ for all $i$*

These conditions, while abstract, are met in many practical cases. For example, the standard model in Equation 1 has $d = 1$, $\boldsymbol{A}_1 = \nu\boldsymbol{1}$ with $\nu > 0$, and $\boldsymbol{\Psi} = \sigma^2\boldsymbol{I}$. Thus $\boldsymbol{A}_1'\boldsymbol{\Psi}_1^{-1}\boldsymbol{A}_1$ is the scalar $(\nu^2/\sigma^2)T \to \infty$ as $T \to \infty$, and the rowsums of $\boldsymbol{\Psi}_1^{-1/2}$ are identically equal to $1/\sigma$. Thus the conditions of the theorem are met, reiterating the results presented in Section 2 that indicated that the random effects estimator converged to the bias-removing fixed-effects estimator as $T$ grew. We provide a few additional examples in which the conditions are met in the simplified case where $\boldsymbol{\Psi} = \sigma^2\boldsymbol{I}$ with $\sigma > 0$, which is representative of more general cases of $\boldsymbol{\Psi} = diag(\sigma_1^2, \ldots, \sigma_T^2)$ with $0 < \sigma_{lower}^2 \leq \sigma_t^2 \leq \sigma_{upper}^2 < \infty$) (both of which satisfy condition 2 of the Theorem). Cases with more general $\boldsymbol{\Psi}_1$ are generally analytically intractable.

**Example 1.** Suppose we generalize the standard individual effects model in Equation 1 to allow the individual effect to be weighted differently for each measurement. In this case, $d = 1$ and $\boldsymbol{A}_1$ is the vector $(a_1, \ldots, a_T)'$. Then $\boldsymbol{A}_1' \boldsymbol{\Psi}_1^{-1} \boldsymbol{A}_1 = (1/\sigma^2) \sum_{t=1}^{T} a_t^2$, and so condition 1 of the theorem is met as long as the series $\sum_{t=1}^{T} a_t^2$ is divergent. A sufficient condition for this divergence is $a_t$ remain bounded away from zero as $T \to \infty$; that is, as long as each observed measurement provides an amount of information about $\delta$ that is not diminishing to zero.

**Example 2.** Suppose $d = 2$ and row $t + 1$ of $\boldsymbol{A}_1$ is $\frac{1}{T-1}(T - 1 - t, t)$ for $t = 0, \ldots, T - 1$. That is, the sequence of measurements gradually changes from all weight on the first factor to all weight on the second factor. Then $\boldsymbol{A}_1' \boldsymbol{\Psi}_1^{-1} \boldsymbol{A}_1 = \frac{1}{\sigma^2 (T-1)^2} \boldsymbol{C}$ where $c_{11} = c_{22} = \frac{1}{3} T^3 + o(T^3)$ and $c_{21} = c_{12} = \frac{1}{6} T^3 + o(T^3)$. The eigenvalues of $C$ can be shown to be $c_{11} \pm c_{21}$, with the smaller eigenvalue thus being $\frac{1}{6} T^3 + o(T^3)$. Multiplying by $\frac{1}{\sigma^2 (T-1)^2}$ gives that the smallest eigenvalue of $\boldsymbol{A}_1' \boldsymbol{\Psi}_1^{-1} \boldsymbol{A}_1$ behaves like $\frac{1}{6\sigma^2} T \to \infty$ as $T \to \infty$.

**Example 3** (Random linear growth model)**.** Suppose $d = 2$ and suppose that the columns of $\boldsymbol{A}_1$ express a random growth model parameterized such that the first column is $(1, 1, \ldots, 1)'$, and the second is $(1, 2, \ldots, T)'$. Then $\boldsymbol{A}_1' \boldsymbol{\Psi}_1^{-1} \boldsymbol{A}_1$ is $(T/\sigma^2)$ times a matrix for which the smallest eigenvalue is bounded away from zero as $T \to \infty$, and thus $\boldsymbol{A}_1' \boldsymbol{\Psi}_1^{-1} \boldsymbol{A}_1$ satisfies the conditions of the theorem.

**Example 4.** Suppose any $d \geq 1$ and suppose that for each measurement the rows of $\boldsymbol{A}$ are a random draw from Dirichlet distribution with parameter $\boldsymbol{\omega}$, a $d \times 1$ vector such that for $j = 1 \ldots d$ $\omega_j > 0$ and $\omega_0 = \sum_{j=1}^{d} \omega_j$. Then $\frac{1}{T} \boldsymbol{A}_1' \boldsymbol{\Psi}_1^{-1} \boldsymbol{A}_1 \to \frac{1}{\sigma^2} (\boldsymbol{\Omega} + c \boldsymbol{\omega} \boldsymbol{\omega}')$, where $\boldsymbol{\Omega}$ is a diagonal matrix with elements $\frac{\omega_0 \omega_j}{\omega_0^2 (\omega_0 + 1)}$ and $c = \frac{\omega_0 (\omega_0 + 1) - 1}{\omega_0^2 (\omega_0 + 1)} > 0$. Theorem 3 on p. 116 of Bellman (7) states that if $\boldsymbol{B}_1$ and $\boldsymbol{B}_2$ are symmetric with $\boldsymbol{B}_2$ is positive semi-definite, then the smallest eigenvalue of $\boldsymbol{B}_1 + \boldsymbol{B}_2$ is greater than or equal to the smallest eigenvalue of $\boldsymbol{B}_1$. Letting $\boldsymbol{B}_1 = \boldsymbol{\Omega}$ and $\boldsymbol{B}_2 = c \boldsymbol{\omega} \boldsymbol{\omega}'$ gives that the smallest eigenvalue of $\frac{1}{\sigma^2} (\boldsymbol{\Omega} + c \boldsymbol{\omega} \boldsymbol{\omega}')$ is greater than or equal to the smallest diagonal element of $\frac{1}{\sigma^2} \boldsymbol{\Omega}$ which is greater than zero. Hence, the eigenvalues of $\boldsymbol{A}_1' \boldsymbol{\Psi}_1^{-1} \boldsymbol{A}_1$ will diverge.

In practical terms, the theorem suggests that when many test scores are available for individual students, the mixed model estimator may be effective at mitigating bias, even if the individual heterogeneity is related to the predictors in the model and even if the covariance structure of the heterogeneity is more complex than the simple time-invariant constant offset model of Equation 1. However, generally characterizing the circumstances under which the elements of $\boldsymbol{R}^{-1} \boldsymbol{A}$ going to zero implies that the mixed model estimator will compress bias as $T \to \infty$ is complex because it depends on the structure of $\boldsymbol{Z}$. A particular challenge is that in most circumstances, growing $T$ implies a growing number of predictors in the model (i.e. columns of $\boldsymbol{Z}$), and in many cases those added predictors apply to only a single time point (e.g time point means or interactions of other variables with time).

One sufficient condition to ensure bias compression as $T \to \infty$ is that the sums of the absolute values of the rows of $\left(\boldsymbol{Z}' \boldsymbol{R}^{-1} \boldsymbol{Z}\right)^{-1} \boldsymbol{Z}'$ are uniformly bounded by a constant (not depending on $T$) as $T \to \infty$. This is not a particularly intuitive condition, and the asymptotic result might not be well approximated for realistic values of $T$. Hence, we now consider several simulation examples building on the examples above to understand when the condition seems likely to hold and when it does not. We consider the behavior of the mixed model estimator to demonstrate the importance of the theorem for reducing bias in estimated effects in situations similar to some applied settings in terms of design, the number of students $n$, the strength of correlation of the test scores within students, and $T$. In the examples, we include the OLS estimator to calibrate the strength of the selection bias.

## 4. Simulation Examples

In this section we consider a series of three simulation examples, building on Examples 1 to 3 in Section 3, to understand the implication of the theorem in real applications. The residual errors of the data generating model of each example ($\boldsymbol{A}\boldsymbol{\delta} + \boldsymbol{\epsilon}$), satisfy the conditions of the theorem. For each example, we consider alternative values for the "treatment" variables $\boldsymbol{Z}$ to understand applications where the theorem will be sufficient for mixed models to remove bias from non-random assignment and where it will not. We also consider different non-random assignment mechanisms that relate components of $\boldsymbol{\delta}$ to $\boldsymbol{Z}$. For reference, Table 1 summarizes the examples that we consider in this section in terms of the model for student heterogeneity, the treatment assignment mechanism, and the configuration of the treatment variables. Additional results from these and other examples, including comparisons to fixed effects estimators, are available in Lockwood and McCaffrey (28).

### *4.1. Example 1, continued*

The model for residual errors in this example is the same as it was in Example 1 of Section 3. Data for 1000 students are generated from a model where each student has a single student effect that is weighted differentially by each measure. The model used to generate the data is:

$$Y_{it} = \boldsymbol{z}_{it}'\boldsymbol{\beta} + a_t\delta_i + \epsilon_{it}$$

where $Y_{it}$ is student $i$'s score in time period $t$ for $t = 1 \ldots T$, $\boldsymbol{z}_{it}$ is a vector of independent variables that can depend on the time period and varies with different scenarios we consider (described below), $\delta_i$ is a random normal variable with mean zero and variance one, $a_t$ is a weight that varies across time, and the $\epsilon_{it}$ are independent random normal variables with mean zero and variance $1-a_t^2$. Thus the variance of the residuals after accounting for treatment effects is equal to 1 for all time points. The $a_t$ vary from 0.7 to 0.9 as $t$ increases from 1 to $T$, so that the correlations of the residuals within students over time are bounded between about 0.5 and 0.8, values consistent with real longitudinal achievement data. If the $a_t$ were constant then the model would be the standard student effect model, so this example is a slight deviation from that case.

| Student Heterogeneity Model | Treatment Assignment | Treatment Scenarios |
|---|---|---|
| **Example 1**<br>Single factor with varying weights over time | Depends on factor | Four scenarios:<br>1: Single Tx effect, students remain in same Tx condition at all time points<br>2: Single Tx effect, students' Tx conditions vary across time points<br>3: Time-varying Tx effects, students remain in same Tx condition at all time points<br>4: Time-varying Tx effects, students' Tx conditions vary across time points |
| **Example 2**<br>Two correlated factors with weights gradually changing from one to the other over time | Three scenarios:<br>1: Tx assignment depends equally on both factors<br>2: Tx assignment depends on only first factor<br>3: Tx assignment depends on only second factor | Treatment only in final year |
| **Example 3 (Teacher Effects)**<br>Linear growth model with positively correlated random slopes and intercepts | Assignment to teachers depends on both intercepts and slopes | Consider one to four tested subjects per year and four different models for teacher effect persistence |

TABLE 1
*Summaries of student heterogeneity model, treatment assignment mechanism, and treatment exposure scenarios for the simulation examples.*

To explore how $\boldsymbol{Z}$ affects the implications of the theorem, we consider four scenarios for the $\boldsymbol{z}_{it}$. All four involve dichotomous treatment indicators where the log odds of treatment assignment equal $\delta_i$ - that is, student assignment to treatment is non-random. In scenario 1, a student's treatment assignment remains the same for all values of $t$ and there is a single treatment effect that is constant for all values of $t$. In scenario 2, a student's treatment assignment can vary across time periods, but there is again only a single treatment effect that is constant for all values of $t$. In scenario 3, a student's treatment assignment remains the same for all values of $t$, but the mean and the treatment effect vary with $t$. In scenario 4, a student's treatment assignment can vary across time periods and the mean and the treatment effect vary with $t$. For each scenario, we consider estimators for $T$ =3, 5, 10, 15, and 20 with 100 Monte Carlo iterations run at each value of $T$.

Figure 1 presents a standardized measure of absolute bias for each estimator (OLS and mixed model) as a function of the scenario and the number of scores. For the mixed model estimator, we use the known value of $\boldsymbol{R}$; we consider the behavior of the estimator when $\boldsymbol{R}$ also must be estimated in Example 2. The standardized bias measure for a given estimator, scenario and number of scores is calculated as follows. Each Monte Carlo simulation provides estimates of the treatment effects (of which there are $T$ for scenarios 3 and 4), the absolute values of which are then divided by the marginal standard deviation in the scores. This makes the metric interpretable in terms of standardized effect sizes. For scenarios 3 and 4, these standardized quantities are then averaged over the multiple treatment effects of these scenarios. Finally the resulting quantities are averaged across Monte Carlo simulations.

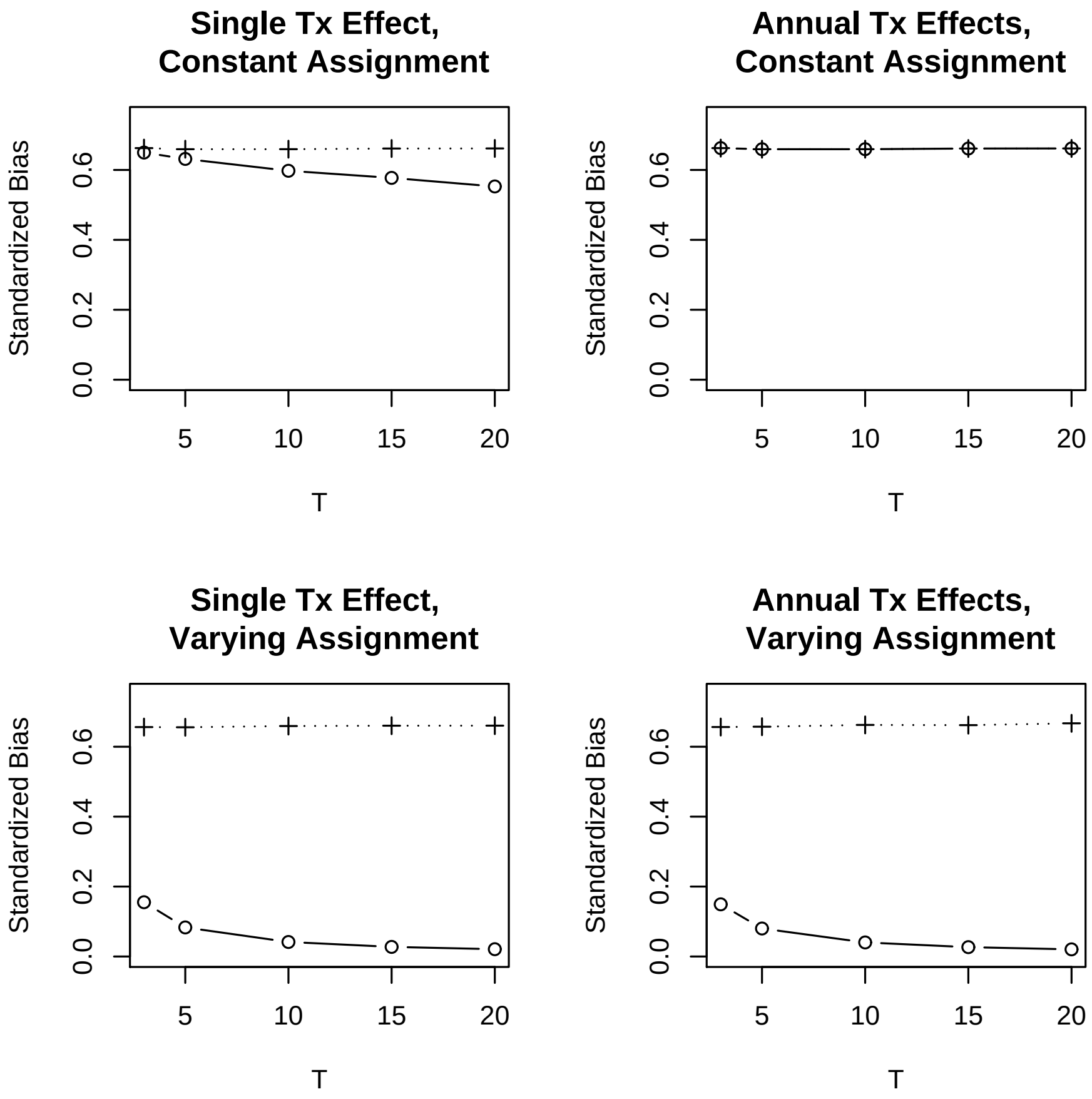

Fig 1. *(Example 1) Average absolute standardized bias as a function of $T$. Different scenarios are shown in different frames. Each frame contains the average absolute standardized bias for OLS (plus signs) and mixed model (open circles).*

As shown in the figure, the mixed model estimator greatly reduces the bias relative to OLS in scenarios 2 and 4 in which treatment assignment varies over time. The bias decreases as $T$ increases in accordance with the theorem. The result holds in scenario 4 even though the number of treatment effects is growing with $T$, indicating that the bias compression is acting effectively across the entire suite of treatment effects. However, when treatment assignment is constant (scenarios 1 and 3; top row of Figure 1), the mixed model estimators are biased and the bias does not decrease with $T$. In these scenarios, the mixed model estimator is similar to the OLS estimator and performs about the same. The row sums of the absolute values of the elements of the matrix $(\boldsymbol{Z}'\boldsymbol{R}^{-1}\boldsymbol{Z})^{-1}\boldsymbol{Z}'$ appear to grow with $T$ when the treatment assignment is constant but not when it is not. The reason for the difference between the estimators is that when treatment assignment is constant each student's contribution to $\boldsymbol{Z}'\boldsymbol{R}^{-1}\boldsymbol{Z}$ is not growing in $T$ because of cancellation between the negative and positive elements of $\boldsymbol{R}^{-1}$. When treatment assignment varies, cancellation does not occur and the elements of $\boldsymbol{Z}'\boldsymbol{R}^{-1}\boldsymbol{Z}$ grow with $T$. This behavior alternatively can be viewed in terms of available information to estimate $\delta_i$ - only when treatment assignment varies can the data provide the information about $\delta_i$ necessary to correct for selection. We conjecture that as long as both the $a_t$ and the overall fraction of non-treatment scores remain bounded away from zero as $T \to \infty$, the bounded rowsums condition will be met and the estimator will be consistent.

### *4.2. Example 2, continued*

This example uses the model for residual errors introduced in Example 2 of Section 3 in which $\delta$ has two components and the contributions of these components to scores gradually changes across tests from one factor to the other. Such a situation might occur if math tests measure two constructs such as computation and problem solving, but the weight given to the two constructs varies systematically across tests. For example, tests for elementary students might focus more on computations whereas tests for middle school students might focus more on problem solving.

The $\boldsymbol{Z}$ used in this model was motivated by a case that is common is applications where $T-1$ test scores are collected on students prior to treatment and then a fraction of the students receive treatment with treatment assignment depending on unobserved characteristics of the student. It is clear that this $\boldsymbol{Z}$ will meet the sufficient condition on the $(\boldsymbol{Z}'\boldsymbol{R}\boldsymbol{Z})^{-1}\boldsymbol{Z}'$ and we showed in Section 3 that the assumptions of the theorem are met. Hence, for large $T$, the treatment effect estimated from the mixed models estimator should be approximately unbiased. The motivation for this example is to explore the properties of the estimator when $T$ is small, to explore the impact of the treatment assignment mechanism on bias, and to explore the impact of estimating $\boldsymbol{R}$ on the performance of the mixed model estimator.

The score $Y_{it}$ for student $i$ on test $t = 1, \ldots T$ is generated according to the model

$$Y_{it} = \mu_t + z_{it}\beta + a_{1t}\delta_{i1} + a_{2t}\delta_{i2} + \epsilon_{it}.$$

The $\mu_t$ are the mean scores for all students by year, and $\beta$ is the effect of treatment and is the parameter of interest. The variable $z_{it}$ is a treatment assignment indicator, and because all but the final test are pre-treatment it is identically equal to zero for all $t < T$. For the roughly one half of the students who receive treatment $z_{iT} = 1$ and for all other students it is zero. The weights $a_{1t}$ are an evenly spaced sequence of values from .1 to .9 and the weights $a_{2t}$ are an evenly spaced sequence of values from .9 to .1. The $\delta_i$ are bivariate normal random variables, with mean zero, variance one and correlation .5. The $\epsilon_{it}$ are i.i.d. $N(0, .2)$ variables. These values imply that correlations among observations from the same student vary from about 0.5 to 0.8 with mean around 0.75.

Using this model, for a sequence of $T$s ranging from 2 to 20, we generated 100 Monte Carlo samples of 1,000 students independently under three scenarios for treatment assignment. In the first scenario, treatment assignment depends equally on $\delta_{i1}$ and $\delta_{i2}$ with the log of the odds of treatment equal to $.4\delta_{i1} + .4\delta_{i2}$. In the second scenario, treatment assignment depends only on $\delta_{i1}$ with the log of the odds of treatment equal to $.4\delta_{i1}$ and in the third scenario, treatment assignment depends only on $\delta_{i2}$ with the log of the odds of treatment equal to $.4\delta_{i2}$. For each scenario, we consider the bias in the estimated the treatment effect using OLS and the mixed model estimator using both the known $\boldsymbol{R}$ as well as $\boldsymbol{R}$ estimated from the data using restricted maximum likelihood implemented in SAS PROC MIXED. We also repeated the estimation using 5,000 observations. The results are extremely similar to the cases with 1,000 and are not presented.

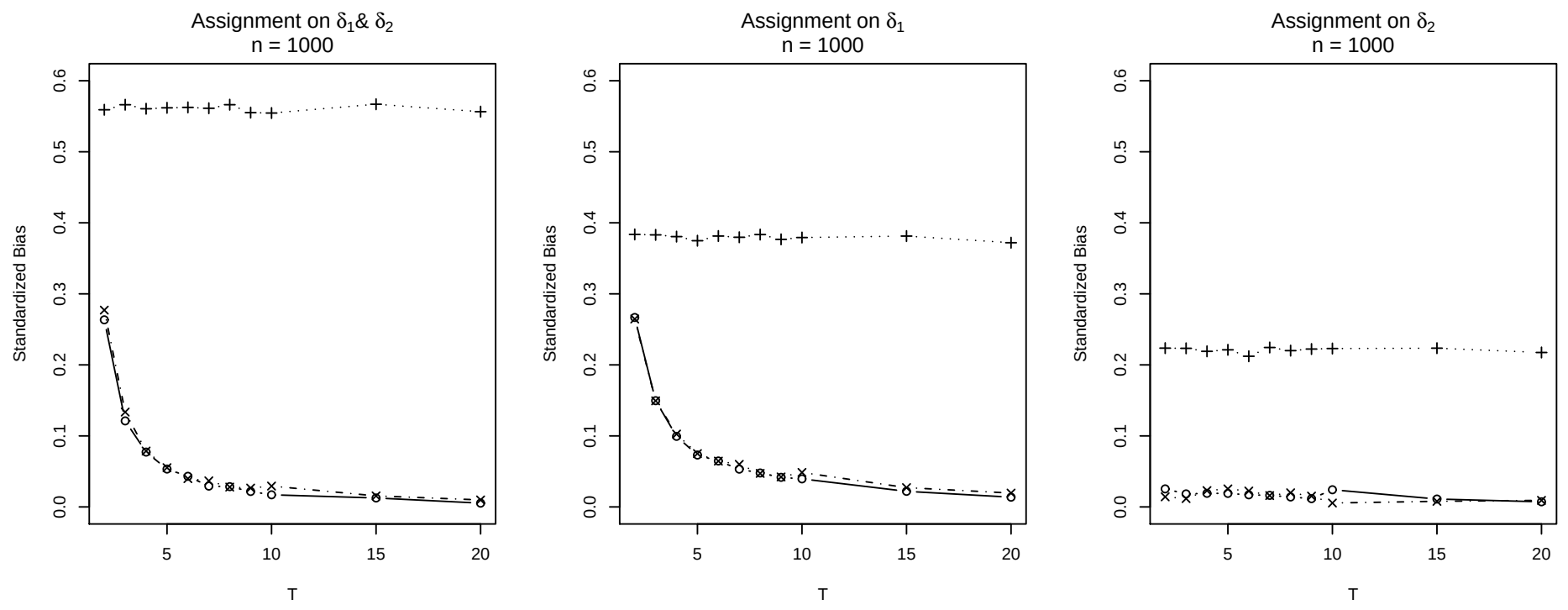

Fig 2. *(Example 2) Standardized absolute bias as a function of $T$. Scenarios 1 to 3 are presented in the panel going from left to right. Each frame contains the average absolute standardized bias for OLS (plus signs), mixed model with known $\boldsymbol{R}$ (open circles), and mixed model with estimated $\boldsymbol{R}$ (X's).*

Figure 2 gives the absolute bias in the estimated treatment effect standardized by the marginal standard deviation in scores for the last period for the three estimators. For all three scenarios, students assigned to treatment have significantly higher $\delta$ values and the OLS estimator has substantial bias, es-pecially in scenarios 1 and 2. For scenario 1, in which treatment assignment depends equally on both elements of $\delta$, the mixed model estimator has rapidly decreasing bias as a function of $T$ and the bias is near zero for $T$ greater than 10. In scenario 2, treatment assignment only depends on $\delta_1$, and again the mixed model estimator has bias approaching zero as $T$ grows, but it declines more slowly than scenario 1 because more recent tests more heavily weight $\delta_1$. In contrast, treatment assignment in scenario 3 depends only on $\delta_2$ which has greater weight in the early scores than the later scores. The OLS estimator using only the test score from the final period in which $\delta_2$ is severely downweighted leads the bias in OLS to be smaller than in the other two scenarios. Similarly, the small weights on $\delta_2$ in final period result in very low bias for the mixed effects estimator for all values of $T$.

In all three scenarios and at all values of $T$, the performance of the mixed model estimator with known $\boldsymbol{R}$ is very similar to the performance of the estimator with $\boldsymbol{R}$ estimated from the data. In scenarios 1 and 2, there is a trend for performance of the estimator with known $\boldsymbol{R}$ to improve relative to the estimator with an estimated $\boldsymbol{R}$ at larger values of $T$ but the differences between the estimators are always less than an percentage point or two. The difference for larger values of $T$ might be due to imprecision in estimating such a large covariance matrix but the similarity at $T = 20$ is very encouraging because the covariance matrix has 210 parameters, but even with as few as 1000 students it estimated with sufficient precision to effectively mitigate the bias.

### *4.3. Example 3, continued: Teacher Effect Estimation*

One of the primary motivating examples for this article is how best to model student heterogeneity when estimating individual teacher effects, as policy interests in using such estimates to reward or sanction teachers is intensifying. The estimation of teacher effects using multivariate longitudinal models is extremely difficult to treat analytically. In the simplest case without any other covariates in the model, the columns of $\boldsymbol{Z}$ correspond to effects of individual teachers on a particular subject, and the elements of the matrix link each student test score to the current and past teachers effects that may have contributed to it. The complexity of $\boldsymbol{Z}$ arises from the crossing of students with teachers as students progress through grades, the assumed model for the persistence of past teacher effects into future test administrations, and the fact that the number of columns of $\boldsymbol{Z}$ grows as more grades and/or subjects are considered. Thus we examine how the mixed model approach behaves through a sequence of simulations.

We consider $n = 1000$ students followed for a period of $G$ consecutive grades, with achievement measured once per grade on $S$ different academic subjects (e.g. reading, mathematics, science, etc). Thus students have a total of $T = GS$ scores. The students are assigned to teachers in each grade, and those teachers are linked to all of the subject scores in each grade. We assume that each class has 25 students, so in this case there are $n/25 = 40$ teachers per grade and thus $40G$ teachers in total across all grades. A separate effect is estimated for each teacher for each subject, so that the total number of teacher effects being estimated is $40GS$.

To simplify the evaluation of the simulation, we generate the data in such a way that there are truly no teacher effects. The score $Y_{isg}$ for student $i$ on subject $s$ in grade $g$ (indexed from 0 to $(G-1)$) is generated according to the random growth model

$$Y_{isg} = \delta_i + \delta_{is} + (\lambda_i + \lambda_{is})g + \epsilon_{isg}$$

similar to that presented in Example 3 in Section 3, except it applies to multiple subjects and the trajectories from different subjects are correlated through the shared intercept parameter $\delta_i$ and shared slope parameter $\lambda_i$. We assume the following independent distributions for the parameters, where all parameters are also independent across students:

$$\begin{aligned} (\delta_i, \lambda_i) &\sim N(\mathbf{0}, \mathbf{\Sigma}) \text{ with } \Sigma_{11} = \sigma_\delta^2, \Sigma_{22} = \sigma_\lambda^2, \Sigma_{21} = \Sigma_{12} = r\sigma_\delta\sigma_\lambda \\ \delta_{i1} \ldots \delta_{iS} &\sim \text{ iid } N(0, \nu_\delta^2) \\ \lambda_{i1} \ldots \lambda_{iS} &\sim \text{ iid } N(0, \nu_\lambda^2) \\ \epsilon_{isg} &\sim \text{ iid } N(0, \sigma_\epsilon^2) \end{aligned} \tag{11}$$

The only parameters that are allowed to be correlated are the common intercept and slope parameters, which have correlation $r$. Under this model, the marginal variance of the scores is the same for all subjects in a given grade, and is $(\sigma_\delta^2 + \nu_\delta^2) + g^2(\sigma_\lambda^2 + \nu_\lambda^2) + 2gr\sigma_\delta\sigma_\lambda + \sigma_\epsilon^2$ for $g = 0, \ldots, (G-1)$. In general scores are correlated across both subjects and grades, with covariances determined by the parameters and the time lags.

Students are regrouped into classes each grade. We introduce spurious teacher effects by making these assignments non-random, and in particular making assignments dependent on the parameters $\delta_i$ and $\lambda_i$. For each student in each grade, we calculate the quantity

$$\eta_{ig} = 0.3(\delta_i/\sigma_\delta) + 0.3(\lambda_i/\sigma_\lambda) + 0.4\xi_{ig}$$

where $\xi_{ig}$ are independent standard normal variables. For each grade, we assign the smallest 25 $\eta_{ig}$ to class 1, the next smallest 25 $\eta_{ig}$ to class 2, etc, all the way to the largest 25 $\eta_{ig}$ to class 40. This results in selection into classrooms that is moderately strong on both student intercepts and student slopes.

In order to estimate teacher effects from these data (the true values of which are identically zero), we need to make assumptions about the design matrix $\boldsymbol{Z}$ that links teacher effect indicators given by the $40GS$ columns to sequences of test scores for students. We assume in all cases that the teacher effect for a given subject affects only the test scores (potentially current and future) for that same subject. For each subject, we assume that the effect of a teacher experienced in grade $g_1$ persists into grade $g_2 \geq g_1$ by the amount $\alpha^{g_2-g_1}$ for some $0 \leq \alpha \leq 1$. The parameter $\alpha$ is taken to be the same for all subjects. The case $\alpha = 0$ corresponds to no persistence of past teacher effects, the case $\alpha = 1$ corresponds to complete persistence of past teacher effects (i.e. the assumption made by the Tennessee Value Added Assessment System (TVAAS) "layered model" of Sanders et al. (41)), and $0 < \alpha < 1$ corresponds to decaying persistence depending on the lag. The design matrices $\boldsymbol{Z}_i$ are fully determined given the sequence of class assignments and $\alpha$.

Our simulation uses $G = 5$, and then varies the number of subjects per grade from 1 to 4 and also considers values of $\alpha$ of 0, 0.3, 0.7 and 1, for a total of 16 design points. We use a common set of variance components and selection model parameters across all design points. In particular we set $\sigma_\delta^2 = 0.5$, $\sigma_\lambda^2 = 0.125$, $r = 0.3$, $\nu_\delta^2 = 0.2$, $\nu_\lambda^2 = 0.05$, $\sigma_\epsilon^2 = 0.8$, which leads to $\boldsymbol{R}$ with marginal variances of 1.500, 1.825, 2.500, 3.525, and 4.900 by grade for each subject, and correlations (both within and across subjects) ranging from about 0.3 to 0.8 with an average of 0.5.

For each design point, we independently generate 100 datasets, and for each dataset we consider three different estimators for the teachers effects: completely unadjusted classroom means, the OLS estimator, and the mixed model estimator using the known value of $\boldsymbol{R}$. The unadjusted classroom means are provided as a strawman to calibrate the strength of selection of students to teachers, like the OLS estimators in the other Examples. For each estimator, rather than summarize the bias in each individual teacher effect, we report estimated variance components of teachers by grade, expressed as a percentage of the corresponding marginal variance in that grade. This standardized measure of estimated teacher variability is commonly used in the literature to summarize the heterogeneity of teacher effects on the scale of student measurements (1; 25; 33; 36; 39). Because our data generating model contains no true teacher effects, the correct value is zero and percentages close to zero indicate that an estimator is behaving effectively. Because of the simplified balanced design of our simulations, the behavior of the estimators is exchangeable across subjects for a given total number of subjects and thus we average the estimated variance components across subjects within grade.

The results are summarized in Figure 3. Each frame of each plot corresponds to a different value of $\alpha$ and presents the estimated teacher variance components by grade and for numbers of subjects of 1, 2, 3 and 4. The lines connect the estimates for a given number of subjects, with the dotted lines corresponding to unadjusted means, dotdash lines corresponding to OLS, and solid lines corresponding to the mixed model estimator. The bias in the unadjusted means and OLS estimators are (up to Monte Carlo error) invariant to the number of subjects, while the bias in the mixed model estimator decreases as the number of subjects increases. As such, the lowest mixed model trajectory in each frame corresponds to 4 subjects, and the highest corresponds to 1 subject.

The unadjusted means indicate that the spurious variation among teachers increases across grades, which makes sense because students are selected into classrooms partially on the basis of growth. When $\alpha = 0$, OLS is equivalent to the unadjusted means but diverges from it as $\alpha$ increases. The decreasing bias in OLS as $\alpha$ increases for grades after the first grade is because the OLS estimator approaches a first-differenced (gain score) estimator as $\alpha$ goes to one.

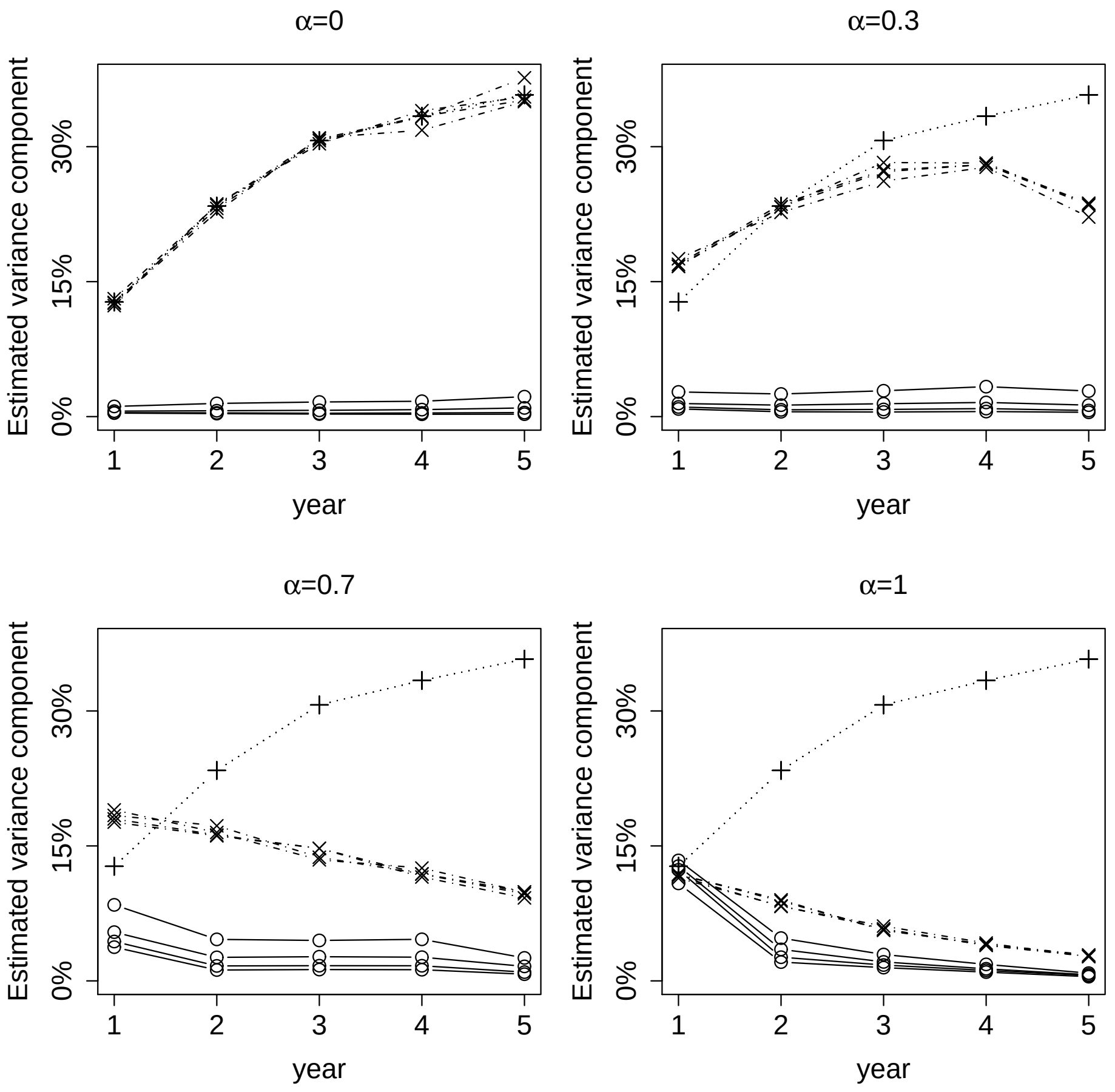

Fig 3. *(Example 3, teacher effects). Estimated teacher variance components, expressed as a fraction of the marginal variance by grade, for unadjusted classroom means (plus signs), OLS (X's) and mixed model (open circles). Each frame of the plot corresponds to a different value of $\alpha$. Multiple lines of the same type within frame correspond to numbers of subjects of 1, 2, 3 and 4 as described in the text.*

This would be sufficient for OLS to remove all bias for teachers beyond the first grade if students were selected into classes on the basis of intercepts alone; however the selection on growth ensures that OLS remains biased. Alternatively, the mixed model estimator is generally effective at removing bias, particularly when $\alpha$ is small, when the bias is uniformly small and effectively zero when multiple subjects are available. As $\alpha$ grows the bias for the mixed model estimator remains for first grade teachers, and when $\alpha = 1$, the mixed model estimator is inconsistent. This is analogous to the cases considered in Example 1 where treatment assignment was constant for all time points, because when $\alpha = 1$ the first grade teacher effect is similarly constant for all time points. Thus in value added studies using mixed model estimators, it is customary not to report estimated teacher effects from the first grade of available teacher links (6; 30)

## 5. Discussion

The primary criticism of random effects or mixed model approaches to analyzing longitudinal data is that they will provide inconsistent estimates when unobserved individual effects are correlated with other variables in the model. Under the standard unobserved effects model, a fixed effects approach does not suffer these same shortcomings and so is widely cited as the preferable analysis technique. However, the results of the econometrics literature indicate that under the standard unobserved effects model, fixed and random effects estimators are very similar when individual heterogeneity accounts for much of the unexplained variation in outcomes and when many measurements are available for each individual. That is, although the random effects estimator can be inconsistent, the magnitude of its bias in practical applications is not necessarily large and the resulting estimates not necessarily poor. The main contribution of this paper is to demonstrate that this "bias compression" property of the mixed model approach extends beyond the standard unobserved effects model to a more general model that might be appropriate for the kinds of complex outcomes considered in social science research, and in particular may be appropriate for analyses of longitudinal student achievement data. This result, in conjunction with the well-established results about the circumstances under which mixed model estimators can be optimally efficient (Aitken's Theorem; see, e.g., Theil (47)), suggests that the mixed model approach may have benefits in longitudinal data analyses that are not widely appreciated.

Intuitively, the mixed model estimator can mitigate bias because pre-multiplication of measurements by $\boldsymbol{R}^{-1}$ serves as a type of regression adjustment. By the results for inverting partitioned matrices (44), pre-multiplication of the measurement vector $\boldsymbol{y}$ by $\boldsymbol{R}^{-1}$ results in values in which each score $y_{it}$ is replaced by a scaled version of the residual $y_{it} - \hat{y}_{it}$ where $\hat{y}_{it}$ is based on the regression of $y_{it}$ on all of the other available measurements for the individual. This adjustment in some sense generalizes the adjustment made by the fixed effects estimator, which replaces each score by its deviation from the average score for each student. Differencing removes all bias due to individual heterogeneity under the standard unobserved effects model, but generally would not be appropriate under the model in Equation 9. Under this general model, as long as individual heterogeneity can be adequately represented by a low-dimensional factor, and the signal about that heterogeneity is not swamped by the noise in the measured outcomes, then the residuals from regression adjustment can be approximately unrelated to the unobserved individual effects, and estimates of model variables can be approximately unbiased. The degree of bias compression improves as both the number of available outcomes, and the signal to noise ratio of those outcomes, increases.

Importantly, our simulation examples demonstrate that the bias compression can hold simultaneously over a suite of parameters whose dimension may be growing as the number of measurement grows, such as with the treatment-by-time interactions considered in Example 1 and the individual teacher effects considered in Example 3. Allowing for this kind of generality in the models may be important in the very circumstances in which the mixed model approach might be warranted - when there is concern that the sequence of tests are not measuring the same construct in the same way over time. It also makes the mixed model approach particularly relevant to jointly modeling measurements on different tested subjects. Nearly all longitudinal achievement data series contain test outcomes from multiple subjects in each year, commonly including both math and reading, often science and/or social studies, and sometimes even multiple scores for the same subject taken from different assessments. It is common for analyses of educational variables, particularly those using fixed effects approaches, for each subject to be modeled independently and estimates reported separately for each subject (c.f. (15; 39; 52)). Analysis of the parallel data series from different subjects in a SUR framework (51) is not common practice in educational research. Our analytical and simulation results suggest that jointly modeling scores from multiple subjects increases the information available for controlling for individual heterogeneity, which can lead to more effective bias compression for all estimated treatment-by-subject effects simultaneously. That is, while the usual justification for SUR analysis is increased efficiency, our results indicate that SUR analysis in the mixed model framework may also provide important bias reduction benefits. The benefits of exploiting the redundancy in repeated measures of multiple outcomes is also noted by Thum (48). Jointly modeling multiple subjects also helps to make the larger values of $T$ considered in our simulation examples more tenable. While following students for $T = 20$ years is unrealistic, using repeated measures on a vector of annual measurements from different subjects or different tests of the same subject makes achieving large numbers of test scores feasible, particularly with increased scope and frequency of standardized testing and the rapidly improving capabilities for linking these scores to students over time.

The ability of the mixed model approach to perform well simultaneously for a number of parameters that grows as the number of test scores grows is particularly relevant for estimating individual teacher effects. For example, the TVAAS model (41) as applied in Tennessee and elsewhere estimates separate teacher-by-subject effects, analogous to Example 3 above, using up to 25 scores for individual students (five subjects for five years). William Sanders, the developer of the TVAAS model, has claimed that jointly modeling 25 scores for students, along with other features of the TVAAS approach, is extremely effective at purging student heterogeneity bias from estimated teacher effects (personal communication). The analytical and simulation results presented here largely support that claim. It is worth noting, however, that TVAAS and related methods of McCaffrey et al. (33) and Raudenbush and Bryk (37) have the additional complexity of modeling individual teacher effects as random effects rather than unknown parameters of indicator variables in a regression model. This makes the analytical considerations more difficult, but as discussed in Lockwood and McCaffrey (28), we believe that the essential elements of the bias compression property of the mixed models discussed here carry over to this more complex class of models.

One critical caveat of our findings is that there are circumstances in which the mixed model approach can fail. Our simulation examples showed cases in which the mixed model approach is unable to remove bias, even with increasing amounts of test score data on individual students. All of these circumstances occurred when student treatment status did not vary over time. This is analogous to the problems faced by fixed effects analyses with similar data, because the treatment variable is confounded with the average of the unobserved student effects for treated students. Intuitively it makes sense that using repeated measures on students to control for student heterogeneity generally is not going to be effective under these circumstances because there is no within-student information upon which to gauge an individual student's performance in the absence of treatment.

A related circumstance when random effects and mixed models cannot control for bias due to unobserved student heterogeneity is when the population is stratified as described in McCaffrey et al. (33). A stratified population is one in which there are disjoint groups of students such that students within a group share teachers but students in different groups never share any teachers. For examples, students from different school districts where there is no interdistrict transfer are a stratified population. As discussed in McCaffrey et al. (33), differences in strata means in $\delta$ are not removed by pre-multiplication by $\boldsymbol{R}^{-1}$. This is because the teacher effects from different strata form "treatments" that are constant for students across the entire time period of the data and, as with constant treatment assignments, random effects cannot mitigate bias.

Another potential limitation of our results is that our analytical results, and all of our examples other than Example 2, treated $\boldsymbol{R}$ as known (using the infeasible GLS estimator), whereas $\boldsymbol{R}$ must be estimated from the data in nearly all practical settings. When $T$ scores are being modeled, $\boldsymbol{R}$ has $T(T+1)/2$ unknown parameters if no additional parametric structure is imposed, and unless the number of students is large relative to $T$ it is likely that $\boldsymbol{R}$ may be estimated with substantial error. Moreover there are cases when $\boldsymbol{R}$ may not even be estimated consistently, such as when students do not switch treatment status over time as in Example 1. Our results from Example 2 where we compared the bias compression from the mixed model estimator using both the known $\boldsymbol{R}$ and $\boldsymbol{R}$ estimated from the data warrant cautious optimism, at least about the effect of estimation error in $\boldsymbol{R}$. In that case the relative abilities of these two estimators to compress bias were almost identical, even for $T = 20$, where our estimated $\boldsymbol{R}$ had 210 parameters and only 1000 students were used in the simulation. We obtained similar results in auxiliary simulations to Example 1 (not shown) where we estimated the model parameters, including $\boldsymbol{R}$, using a Bayesian model. We conjecture that for the $\boldsymbol{R}$s that are likely to exist in longitudinal student achievement data series - where a large portion of the residual variance is dominated by a low-dimensional student-specific factor - that estimation error in $\boldsymbol{R}$ is not likely to substantially degrade the bias compression property of the mixed model estimator because the redundant information available in the vector of scores is likely to lead to estimates of $\hat{y}_{it}$ that are largely insensitive to error in the estimated coefficients of the regression adjustment performed by $\boldsymbol{R}^{-1}$. Further exploration of this issue in settings more complex than that considered in Example 2 is an important area for future research.

Also, all of the cases considered here had balanced, completely observed score data for all students. Actual data sets invariably contain a substantial amount of missing test score information, and when multiple cohorts are being modeled, it is common for different cohorts to have different configurations of available test scores. To explore the effects of missing data, we expanded Example 2 by adding cases where 50 percent of the data are missing at random. The bias continues converge to zero with increasing $T$ but the decay is reduced so that the bias with $T = 20$ and 50 percent of observations missing is similar to the bias with $T = 15$ and all the data are observed. This suggests that our general findings about the bias compression of the mixed models approach are not invalidated by the complexities of missing data, but it is likely that incompleteness in the test score data will in general degrade the bias compression to some extent. On the other hand, the mixed models approach makes use of all of the information available for each student in estimating the unknown parameters - in essence estimating $\hat{y}_{it}$ from the regression on the available scores for each student - and so might lead to particular efficiency gains relative to other approaches when missing data are substantial.

In summary, our results indicate that the fact that mixed model approaches to longitudinal data analysis can lead to inconsistent estimates does not mean that the estimates are necessarily poor in any given instance. When applied to longitudinal data involving a large number of correlated measurements on individuals, they can provide nearly unbiased estimates even under relatively complex heterogeneity models involving multiple, unobserved individual-specific attributes whose relationship to the observed measurements varies across those measurements. Importantly, the bias compression is a byproduct of GLS estimation and does not require specification of $\boldsymbol{R}_1$ - for example, it is not necessary to decide if a one-factor or two-factor heterogeneity model is more appropriate. This kind of “black-box” robustness of the mixed model approach might be beneficial in circumstances such as longitudinal achievement data series where the true heterogeneity model may be complex and obtaining formally consistent estimates may require more advanced methods. For example, Ahn, Lee and Schmidt (2) and Han, Orea and Schmidt (18) develop and compare consistent generalized method of moments and concentrated least squares estimators under structural models similar to our Example 1, where the coefficients on the individual-specific factors vary across measurements. Similarly, covariance structure modeling approaches (5; 12; 46) are well-suited to handling cases where factors are related to measurements in time-varying ways. Understanding the performance of the mixed model approach in terms of both bias compression and precision relative to these alternatives warrants further study.

It also important that future work study the performance of the mixed model approach relative to empirical modificiations to the standard fixed effects model that attempt to address some of the complexities discussed in this article. For example, with longitudinal achievement data, analysts often internally standardize each test measure to have constant variance of one or replace scores by ranks or transformed ranks prior to analysis as a strategy to increase the plausibility of the standard unobserved effects model when there are concerns about the comparability of scales across measurements (14; 15; 52). This procedure is appropriate under particular instances of the model in Equation 9 but not in general. Thus standard fixed effects approaches applied to the transformed data are not likely to provide formally consistent estimates, and it would be useful to understand how the degree of inconsistency compares to the mixed model approach. Similarly, it is a common strategy for analysts to use fixed effects on student timepoint-to-timepoint gain scores rather than student level scores when there are concerns about selection bias due to differential grade-to-grade growth in achievement (20; 52). This approach has strict data requirements because multiple year-to-year gain measures must be available on a student in order for that student to contribute to the estimation of the model parameters, which can result in substantial reduction in the amount of usable information given the degree of missing data commonly found. The mixed model estimator does not impose similar restrictions, and so might have lower mean squared error even though it may be inconsistent.

## 6. Appendix A

**Theorem.** *Let $\boldsymbol{A}_1$ and $\boldsymbol{\Psi}_1$ be defined as in Section 3. Then sufficient conditions for the elements of $\boldsymbol{R}^{-1}\boldsymbol{A}$ to go to zero uniformly as $T \to \infty$ are:*

1. *The smallest eigenvalue of $\boldsymbol{A}_1'\boldsymbol{\Psi}_1^{-1}\boldsymbol{A}_1$ goes to infinity as $T \to \infty$; and*
2. *There exists a number $C$ independent of $T$ such that the elements $a_{it}$ of $\boldsymbol{\Psi}_1^{-1/2}$, the symmetric square root of $\boldsymbol{\Psi}_1^{-1}$, satisfy $\sum_{t=1}^{T} |a_{it}| < C$ for all $i$*

*Proof.* Throughout, all matrices except $\boldsymbol{S}_1$ and its root are assumed to depend on $T$ so we suppress that notation. Because $\boldsymbol{R}^{-1}\boldsymbol{A} = (\boldsymbol{I}_n \otimes \boldsymbol{R}_1^{-1})(\boldsymbol{I}_n \otimes \boldsymbol{A}_1) = \boldsymbol{I}_n \otimes \boldsymbol{R}_1^{-1}\boldsymbol{A}_1$, it is sufficient to consider only the elements of $\boldsymbol{R}_1^{-1}\boldsymbol{A}_1$. Because all matrices would thus be subscripted by "1", we suppress that notation as well and use, for example, $\boldsymbol{R}$ for $\boldsymbol{R}_1$, $\boldsymbol{A}$ for $\boldsymbol{A}_1$, $\boldsymbol{\Psi}$ for $\boldsymbol{\Psi}_1$, etc. We also assume that all matrix roots are symmetric roots.

Recall that $\boldsymbol{R} = \boldsymbol{A}\boldsymbol{S}\boldsymbol{A}' + \boldsymbol{\Psi}$. By the Schur complement formula (44)

$$\boldsymbol{R}^{-1} = \boldsymbol{\Psi}^{-1}\left[\boldsymbol{I} - \boldsymbol{A}\boldsymbol{S}^{1/2}(\boldsymbol{I} + \boldsymbol{S}^{1/2}\boldsymbol{A}'\boldsymbol{\Psi}^{-1}\boldsymbol{A}\boldsymbol{S}^{1/2})^{-1}\boldsymbol{S}^{1/2}\boldsymbol{A}'\boldsymbol{\Psi}^{-1}\right]$$

so that

$$\begin{aligned}\boldsymbol{R}^{-1}\boldsymbol{A} &= \boldsymbol{\Psi}^{-1}\boldsymbol{A}\left[\boldsymbol{I} - \boldsymbol{S}^{1/2}(\boldsymbol{I} + \boldsymbol{S}^{1/2}\boldsymbol{A}'\boldsymbol{\Psi}^{-1}\boldsymbol{A}\boldsymbol{S}^{1/2})^{-1}\boldsymbol{S}^{1/2}\boldsymbol{A}'\boldsymbol{\Psi}^{-1}\boldsymbol{A}\right] \\ &= \boldsymbol{\Psi}^{-1}\boldsymbol{A}\boldsymbol{S}^{1/2}\left[\boldsymbol{I} - (\boldsymbol{I} + \boldsymbol{S}^{1/2}\boldsymbol{A}'\boldsymbol{\Psi}^{-1}\boldsymbol{A}\boldsymbol{S}^{1/2})^{-1}\boldsymbol{S}^{1/2}\boldsymbol{A}'\boldsymbol{\Psi}^{-1}\boldsymbol{A}\boldsymbol{S}^{1/2}\right]\boldsymbol{S}^{-1/2}\end{aligned}$$

Let $\boldsymbol{X} = \boldsymbol{\Psi}^{-1/2}\boldsymbol{A}\boldsymbol{S}^{1/2}$ of dimension $(T \times d)$. Then

$$\boldsymbol{R}^{-1}\boldsymbol{A} = \boldsymbol{\Psi}^{-1/2}\boldsymbol{X}\left[\boldsymbol{I} - (\boldsymbol{I} + \boldsymbol{X}'\boldsymbol{X})^{-1}\boldsymbol{X}'\boldsymbol{X}\right]\boldsymbol{S}^{-1/2}$$

Singular value decompose $\boldsymbol{X}$ as $\boldsymbol{U}\boldsymbol{\Lambda}^{1/2}\boldsymbol{V}'$ where $\boldsymbol{U}$ is $(T \times d)$ with orthonormal columns, $\Lambda^{1/2} = diag(\sqrt{\lambda_1}, \ldots, \sqrt{\lambda_d})$, and $\boldsymbol{V}$ is $(d \times d)$ and orthogonal. Because $\boldsymbol{A}$ is assumed to have full column rank, $\boldsymbol{X}$ has full column rank and so $\sqrt{\lambda_m} > 0$ for $m = 1, \ldots, d$. Note that $\boldsymbol{X}'\boldsymbol{X} = \boldsymbol{S}^{1/2}\boldsymbol{A}'\boldsymbol{\Psi}^{-1}\boldsymbol{A}\boldsymbol{S}^{1/2} = \boldsymbol{V}\boldsymbol{\Lambda}\boldsymbol{V}'$ where $\lambda_1, \ldots, \lambda_d$ are the eigenvalues of $\boldsymbol{X}'\boldsymbol{X}$.

Now consider the matrix

$$\begin{aligned}
\left[\boldsymbol{I} - (\boldsymbol{I} + \boldsymbol{X}'\boldsymbol{X})^{-1}\boldsymbol{X}'\boldsymbol{X}\right] &= \left[\boldsymbol{I} - (\boldsymbol{I} + \boldsymbol{V}\boldsymbol{\Lambda}\boldsymbol{V}')^{-1}\boldsymbol{V}\boldsymbol{\Lambda}\boldsymbol{V}'\right] \\
&= \left[\boldsymbol{V}\boldsymbol{V}' - \boldsymbol{V}(\boldsymbol{I} + \boldsymbol{\Lambda})^{-1}\boldsymbol{V}'\boldsymbol{V}\boldsymbol{\Lambda}\boldsymbol{V}'\right] \\
&= \left[\boldsymbol{V}(\boldsymbol{I} - (\boldsymbol{I} + \boldsymbol{\Lambda})^{-1}\boldsymbol{\Lambda})\boldsymbol{V}'\right] \\
&= \left[\boldsymbol{V} diag(\frac{1}{1+\lambda_1}, \ldots, \frac{1}{1+\lambda_d})\boldsymbol{V}'\right]
\end{aligned}$$

Thus

$$\begin{aligned}
\boldsymbol{R}^{-1}\boldsymbol{A} &= \boldsymbol{\Psi}^{-1/2}\boldsymbol{U}\boldsymbol{\Lambda}^{1/2}\boldsymbol{V}'\left[\boldsymbol{V} diag(\frac{1}{1+\lambda_1}, \ldots, \frac{1}{1+\lambda_d})\boldsymbol{V}'\right]\boldsymbol{S}^{-1/2} \\
&= \boldsymbol{\Psi}^{-1/2}\boldsymbol{U}\boldsymbol{\Lambda}^{*}\boldsymbol{V}'\boldsymbol{S}^{-1/2}
\end{aligned}$$

where $\boldsymbol{\Lambda}^*$ is $diag(\sqrt{\lambda_1}/(1+\lambda_1), \ldots, \sqrt{\lambda_d}/(1+\lambda_d))$.

An arbitrary element $r_{ij}$ of $\boldsymbol{R}^{-1}\boldsymbol{A}$ is the inner product of a row $\boldsymbol{a}_i^{'}$ of $\boldsymbol{\Psi}^{-1/2}$ and a column $\boldsymbol{b}_j$ of $\boldsymbol{U}\boldsymbol{\Lambda}^*\boldsymbol{V}\boldsymbol{S}^{-1/2}$. Let $s$ be the absolute value of the largest element of $\boldsymbol{S}^{-1/2}$. By condition 1 of the theorem and Lemma **??**, all the eigenvalues of $\boldsymbol{X}'\boldsymbol{X}$ are getting arbitrarily large for sufficiently large values of $T$, so that for any $\epsilon > 0$ there exists a $T_\epsilon$ such that $\sqrt{\lambda_m}/(1+\lambda_m) < \epsilon/Csd^2$ for all $m = 1, \ldots, d$ and for all $T > T_\epsilon$. Because $\boldsymbol{V}$ is an orthogonal matrix and the columns of $\boldsymbol{U}$ are orthonormal, their elements cannot exceed 1 in absolute value, and so the absolute value of the largest element of $\boldsymbol{U}\boldsymbol{\Lambda}^*\boldsymbol{V}\boldsymbol{S}^{-1/2}$ is bounded by $\epsilon/C$ for all $T > T_\epsilon$. Then, for any $i$ and $j$ and for all $T > T_\epsilon$,

$$|r_{ij}| = |\boldsymbol{a}_i^{'}\boldsymbol{b}_j| = |\sum_{t=1}^{T} a_{i,t}b_{j,t}| \leq \sum_{t=1}^{T} |a_{i,t}||b_{j,t}| \leq (\epsilon/C)\sum_{t=1}^{T} |a_{i,t}| < \epsilon$$

where the last inequality follows by condition 2 of the theorem. □

**Lemma 1.** *Let* $\boldsymbol{B}$ *be a positive definite* $d \times d$ *matrix and let* $\boldsymbol{B}^{1/2}$ *be a symmetric root of* $\boldsymbol{B}$. *Let* $\boldsymbol{M}_T$ *be a sequence of* $d \times d$ *matrices. Let* $\lambda_T$ *denote the smallest eigenvalue of* $\boldsymbol{M}_T$ *and* $\omega_T$ *denote the smallest eigenvalue of* $\boldsymbol{Q}_T = \boldsymbol{B}^{1/2}\boldsymbol{M}_T\boldsymbol{B}^{1/2}$. *Then* $\lambda_T \to \infty$ *as* $T \to \infty$ *if and only if* $\omega_T \to \infty$ *as* $T \to \infty$.

*Proof.* Suppose $\lambda_T \to \infty$ as $T \to \infty$. Let $\psi_{min} > 0$ denote the smallest eigenvalue of $\boldsymbol{B}$. Then for every $d$-vector $\boldsymbol{x}$ such that $\boldsymbol{x}'\boldsymbol{x} = 1$

$$\begin{aligned}\boldsymbol{x}'\boldsymbol{Q}_T\boldsymbol{x} &= (\boldsymbol{x}'\boldsymbol{B}\boldsymbol{x})\frac{\boldsymbol{x}'\boldsymbol{Q}_T\boldsymbol{x}}{\boldsymbol{x}'\boldsymbol{B}\boldsymbol{x}} \\ &\geq \psi_{min}\frac{\boldsymbol{x}'\boldsymbol{Q}_T\boldsymbol{x}}{\boldsymbol{x}'\boldsymbol{B}\boldsymbol{x}} \\ &= \psi_{min}\frac{\boldsymbol{u}'\boldsymbol{M}_T\boldsymbol{u}}{\boldsymbol{u}'\boldsymbol{u}}, \text{ where } \boldsymbol{u} = \boldsymbol{B}^{1/2}\boldsymbol{x} \\ &\geq \psi_{min}\lambda_T.\end{aligned}$$

Because $\boldsymbol{B}$ is positive definite, division by $\boldsymbol{x}'\boldsymbol{B}\boldsymbol{x}$ is well defined. By assumption $\psi_{min}\lambda_T \to \infty$ with $T$ so that $\omega_T$, the minimum of $\boldsymbol{x}'\boldsymbol{Q}_T\boldsymbol{x}$, must converge to infinity as well.

Now suppose that $\omega_T \to \infty$ as $T \to \infty$. Let $\psi_{max} > 0$ denote the largest eigenvalue of $\boldsymbol{B}$. Then for any $d$-vector $\boldsymbol{a}$, $\boldsymbol{a}'\boldsymbol{B}\boldsymbol{a} \leq \psi_{max}\boldsymbol{a}'\boldsymbol{a}$. Now, let $\boldsymbol{u}$ be a vector such that $\lambda_T = (\boldsymbol{u}'\boldsymbol{M}_T\boldsymbol{u})/\boldsymbol{u}'\boldsymbol{u}$ and let $\boldsymbol{x} = \boldsymbol{B}^{-1/2}\boldsymbol{u}$ so that $\boldsymbol{B}^{1/2}\boldsymbol{x} = \boldsymbol{u}$. Then

$$\begin{aligned}\lambda_T &= \frac{\boldsymbol{x}'\boldsymbol{B}^{1/2}\boldsymbol{M}_T\boldsymbol{B}^{1/2}\boldsymbol{x}}{\boldsymbol{x}'\boldsymbol{B}\boldsymbol{x}} \\ &\geq \frac{\boldsymbol{x}'\boldsymbol{B}^{1/2}\boldsymbol{M}_T\boldsymbol{B}^{1/2}\boldsymbol{x}}{\psi_{max}\boldsymbol{x}'\boldsymbol{x}} \\ &\geq \frac{\omega_T}{\psi_{max}} \to \infty \text{ as } T \to \infty.\end{aligned}$$

□